\documentclass[preprint, showpacs,preprintnumbers,amsmath,amssymb,nofootinbib]{revtex4}

\usepackage{epsf}
\usepackage{graphicx}

\newcommand{\be}[1]{\begin{equation}\label{#1}}
 \newcommand{\ee}{\end{equation}}
 \newcommand{\bea}{\begin{eqnarray}}
 \newcommand{\eea}{\end{eqnarray}}

 \def\gsim{ \lower .75ex \hbox{$\sim$} \llap{\raise .27ex \hbox{$>$}} }
 \def\lsim{ \lower .75ex \hbox{$\sim$} \llap{\raise .27ex \hbox{$<$}} }

\begin{document}
 \title{Cosmological evolution of interacting phantom (quintessence) model in Loop Quantum Gravity}

\author{ Puxun Wu\;$^{1,2}$\footnote{wpx0227@gmail.com} and Shuang Nan Zhang\;$^{1,3,4}$}

\address
{$^1$Department of Physics and Tsinghua Center for Astrophysics,
Tsinghua University, Beijing 100084, China
\\
$^2$School  of Sciences and Institute of  Math-Physics, Central
South University of Forestry and Technology, Changsha, Hunan 410004,
China
\\
 $^3$ Key Laboratory of Particle Astrophysics, Institute of
High Energy Physics, Chinese Academy of Sciences, P.O. Box 918-3,
Beijing 100049, China
\\
$^4$ Physics Department, University of Alabama in Huntsville,
Huntsville, AL 35899, USA }

\begin{abstract}
The dynamics of interacting dark energy model in loop quantum
cosmology (LQC) is studied in this paper. The dark energy has a
constant equation of state $w_x$ and interacts with dark matter
through a form  $3cH(\rho_x+\rho_m)$. We find for quintessence model
($w_x>-1$) the cosmological evolution in LQC is the same as that in
classical Einstein cosmology; whereas for phantom dark energy
($w_x<-1$), although there are the same critical points in LQC and
classical Einstein cosmology, loop quantum effect reduces
significantly the parameter spacetime ($c, w_x$) required by
stability. If parameters $c$ and $w_x$ satisfy the conditions that
the critical points are existent and stable, the universe will enter
an era dominated by dark energy and dark matter with a constant
energy ratio between them, and accelerate forever; otherwise it will
enter an oscillatory regime. Comparing our results with the
observations we find at $1\sigma$ confidence level the universe will
accelerate forever.
\end{abstract}


 \maketitle

\section{Introduction}\label{sec1}
Many cosmological  observations show that our universe is undergoing
an accelerating expansion and now mainly consists of  two dark
components: dark matter and dark energy. The dark matter is a clumpy
component that traces the baryonic matter and accounts for about
$23\%$ of present total cosmic energy;  the dark energy is an exotic
energy with negative pressure and accounts for about $72\%$ of total
cosmic energy  today. The simplest candidate of dark energy is the
cosmological constant~\cite{Constant}, however it suffers from two
problems. One is the cosmological constant problem: why is the
inferred value of cosmological constant so tiny (120 orders of
magnitude lower) compared to the typical vacuum energy values
predicted by the quantum field theory? The other is the coincidence
problem: why is the dark energy density comparable to the matter
density right now? Therefore a dynamical scalar field:
quintessence~\cite{Quint} is proposed as an alternative of dark
energy, but it can not explain the region of the equation of state
less than $-1$, which is favored by observations~\cite{Nesseris}.
Later Caldwell~\cite{Cald} proposed a phantom field to explain the
present cosmic accelerating expansion. This field possesses of a
negative kinetic energy and so has a super negative equation of
state. In the Einstein gravity it is found that if the universe is
dominated by the phantom energy, it will end with a big rip, i.e., a
future singularity~\cite{Cald2}. Many works have been done trying to
avoid this singularity~\cite{McInne}. There are many other scalar
field models: such as quintom~\cite{Quintom} and
hessence~\cite{Hessence}. However these scale field dark energy
models still suffer from the coincidence problem. A possible
alleviation for this problem is to assume the existence of an
interaction between dark matter and dark energy~\cite{Chimento2003}.

Recent investigations have shown that there are some new nice
features appearing in Loop Quantum Cosmology (LQC) (see \cite{LQC,
LQC2} for recent reviews), such as: easier inflation~\cite{Easy} and
correspondence between LQC and braneworld
cosmology~\cite{Copeland2006}. The LQC is the application in the
cosmology context of the Loop Quantum Gravity (LQG) (see \cite{LQG}
for recent reviews), which is a theory  trying to quantize the
gravity with a non-perturbative and background-independent method.
Due to the loop quantum effect the standard Friedmann equation can
be modified by adding a correction term~\cite{LQC2, Fried1,
Fried2,Fried3}, \bea
H^2=\frac{1}{3}\rho\bigg(1-\frac{\rho}{\rho_c}\bigg)\;,\eea where
$H$ is the Hubble parameter, $8\pi G\equiv 1$,  $\rho$ is the total
cosmic energy density, $\rho_c\equiv\frac{\sqrt{3}}{16\pi^2\gamma^3
G^2 \hbar}$ denotes the critical loop quantum density  and $\gamma$
is the dimensionless Barbero-Immirzi parameter (it is suggested that
$\gamma=0.2375$ by the black hole thermodynamics in LQG
\cite{Ashtekar}). Since this modified equation is correct under the
condition that the quantum state is semiclassical, this condition is
assumed to be satisfied forever in this paper. In addition we assume
the quantum correlations do not build up during long-term evolution
of cosmology, otherwise there are additional correction terms from
LQC which become important~\cite{Bojowald2008}. The correction term
appearing in Eq.~(1) essentially encodes the discrete quantum
geometric nature of spacetime.  When this correction term becomes
dominant, the universe begins to bounce and then expands backwards.
By studying the early universe inflation and the fate of future
singularity in LQC,  it is found that the big bang singularity, the
big rip  and other future singularities can be avoided~\cite{LQC2,
Fried1, Fried2, Samart}. Recently Samart and
Gumjudpai~\cite{Samart}, Wei and Zhang~\cite{Hao} studied the
dynamics of phantom, quintom and hessence dark energy models in LQC,
and found the results are different from that obtained in classical
Einstein cosmology. In this paper we will investigate the evolution
of our universe dominated by a scalar field in LQC, which has
constant equation of state and interacts with dark matter,   and
then investigate whether there are some interesting features arising
from the loop quantum gravity effect.

\section{The interacting model}
We consider a spatially flat universe in which there are only dark
matter and dark energy with a constant equation of state $w_x$.
Apparently $w_x>-1$ corresponds a quintessence model and $w_x<-1$ is
a phantom case. In addition we assume that between the dark matter
and dark energy there is an interaction term $\Gamma$. Thus the
conservation equations for dark matter and dark energy can be
expressed as \bea\label{rhox}
\dot{\rho}_x+3H(1+w_x)\rho_x=-\Gamma\;,\eea \bea\label{rhom}
\dot{\rho}_m+3H\rho_m=\Gamma\;,\eea where $\rho_x$ and $\rho_m$
correspond to the energy densities of dark energy and dark matter
respectively,  and a dot denotes the derivative with respect to
cosmic time $t$. The interacting term $\Gamma$ is assumed to be
$\Gamma=3Hc(\rho_{x}+\rho_{m})$, where $c$ is a coupling constant
denoting the transfer strength. A positive $c$ corresponds to energy
transferred from dark energy to dark matter and the other way around
for a negative one. In this paper we constrain our discussion in the
case of $c>0$. This type of interaction,  motivated by analogy with
dissipation of cosmological fluids, has been introduced to solve the
coincidence problem~\cite{Chimento2003}, and has been studied in the
context of quintessence~\cite{Zimdahl}, phantom~\cite{Guo} and the
(generalized) Chaplygin gas model~\cite{Zhang}. In addition the
observational constraints for this type interaction dark energy
model have been studied in Refs.~\cite{Olivares, Wang}.

In LQC, using the conservation equation of cosmic total energy $
 \dot{\rho}+3H\left(\rho+p\right)=0$, where $\rho=\rho_x+\rho_m$,
  one can easily obtain the effective modified Raychaudhuri
equation
 \bea\label{ray}
 \dot{H}=-\frac{~1}{~2}\left(\rho+p\right)
 \left(1-2\frac{\rho}{\rho_{\rm c}}\right),
 \eea
 where $p$ is the total pressure ($p=w_x \rho_x$ in this paper).

\begin{figure}[htbp]
\includegraphics[width=10cm]{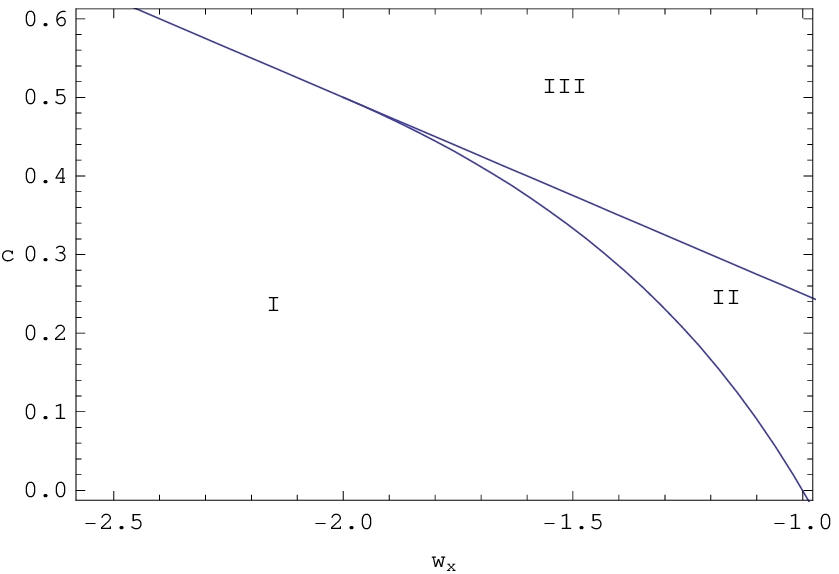}
 \caption{\label{Fig1} The stable regions of ($w_x$, $c$) parameter space.
In the region II, critical point A is a late time attractor in LQC.
In Einstein  cosmology critical point A is a late time attractor in
the region I+II. III represents the region of the solution without
physical meaning.}
\end{figure}

To analyze the dynamical system, we set
 \bea N=\ln a, \qquad u=\frac{\rho_{x}}{3H^2}, \qquad v=\frac{\rho_m}{3H^2}, \eea
where $a_0=1$ is assumed. Using Eqs.(\ref{rhox}, \ref{rhom},
\ref{ray}), one can obtain \bea\label{s1}
u'=-3u(1+w_x)-3c(u+v)+3u(u+w_x u+v)\big(-1+\frac{2}{u+v}\big)\;,\eea
\bea\label{s2} v'=-3v+3c(u+v)+3v(u+w_x
u+v)\big(-1+\frac{2}{u+v}\big)\;, \eea where the prime denotes a
derivative with respect to $N$. As discussed in
Refs.~\cite{Chimento2003, Guo, Olivares} this interacting model can
solve, or at least ameliorate the coincidence problem in classical
Einstein cosmological since in the dynamical system there is a late
time attractor  solution with a constant energy ratio between dark
energy and dark matter. Therefore, regardless the initial
conditions, the universe evolves to a final state characterized by a
constant dark matter to dark energy ratio. Here we will discuss in
LQC whether the dynamics system of interacting model exists the
attractor solution, and then study the cosmic evolutions within
different conditions. In order to obtain the possible attractor for
the system given by Eqs.~(\ref{s1}, \ref{s2}), we should firstly
solve these equations with $u'=0$ and $v'=0$ to get the critical
points:
  \bea Point\; A:
u_c=\frac{1}{2}+\frac{1}{2}\sqrt{1+\frac{4c}{w_x}}, \qquad
v_c=1-u_c. \eea
 \bea
Point\;  B: u_c=\frac{1}{2}-\frac{1}{2}\sqrt{1+\frac{4c}{w_x}},
\qquad v_c=1-u_c\;. \eea Both two critical points correspond to the
era dominated by dark matter and dark energy with a constant energy
ratio between them and exist for $c\leq\frac{-w_x}{4}$. Apparently
these critical points are the same as that obtained in classical
Einstein cosmology~\cite{Olivares}.  If the critical point is
stable, it is a late time attractor; otherwise the solution is
oscillatory. In order to investigate the stability of the critical
point, we linearize the system near the critical point and arrive at
\bea \delta u'&=&-\bigg[3c+3(-1+2u_c+v_c)+3w_x+6w_x
u_c\bigg(1-\frac{u_c+2v_c}{(u_c+v_c)^2}\bigg)\bigg] \delta u
\nonumber \\ &&- \bigg [3(c+u_c)+\frac{6w_x
u_c^2}{(u_c+v_c)^2}\bigg]\delta v\eea
 \bea
\delta
v'&=&\bigg[3c+3(1+w_x)(-v_c+\frac{2v_c}{u_c+v_c})-6\frac{v_c(u_c+w_x
u_c+v_c)}{(u_c+v_c)^2}\bigg] \delta u \nonumber \\ &&- \bigg
[3+3c-3(1+w_x)u_c-6v_c+\frac{6w_x u_c^2}{(u_c+v_c)^2}\bigg]\delta
v\eea Apparently there are two eigenvalues of the coefficient matrix
of the above equations. If the real parts of two eigenvalues for a
critical point are all negative, this critical point is stable and
is an attractor; otherwise it is unstable and thus oscillatory. We
find the point B is always unstable, however the point A is   an
attractor if the equation of state for dark energy $w_x$ and the
coupling factor $c$ satisfy the conditions $\frac{1+w_x}{w_x}\leq
c\leq \frac{-w_x}{4}$ and $w_x>-2$. Comparing our results with that
obtained in Einstein cosmology where point A is stable only under
condition $c\leq\frac{-w_x}{4}$~\cite{Olivares}, we find that for
quintessence dark energy $w_x>-1$ the results in LQC are the same as
that in Einstein  cosmology if a positive $c$ is considered  since
$\frac{1+w_x}{w_x}<0$. However for phantom dark energy $w_x<-1$ the
conclusions seem to be different: in region  $c<\frac{1+w_x}{w_x}$
or $w_x<-2$, the point A is stable in Einstein cosmology, but it is
unstable in LQC,  that is,  the quantum correction effect will break
the stability of point A if the interaction factor $c$ is smaller
than $\frac{1+w_x}{w_x}$ or the equation of state for phantom is
less than $-2$. In Fig.~(\ref{Fig1}) we show the stability regions
of $(c,w_x)$ parameter space.  Regions I$+$II are allowed for
Einstein cosmology; however in LQC only region II is allowed to
obtain a stable solution.

\begin{figure}[htbp]
\includegraphics[width=10cm]{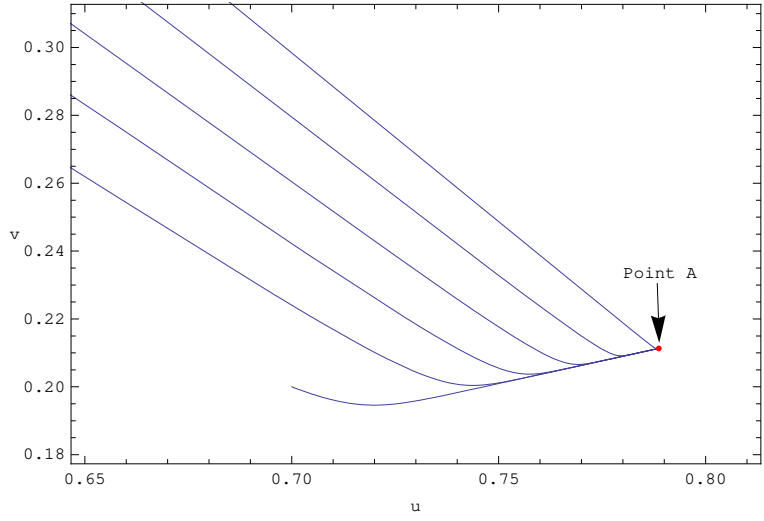}
 \caption{\label{Fig2} The phase diagram of the interacting phantom
 dark energy in LQC  with $w_x=-1.2$ and $c=0.2$.  }
\end{figure}

\begin{figure}[htbp]
\includegraphics[width=10cm]{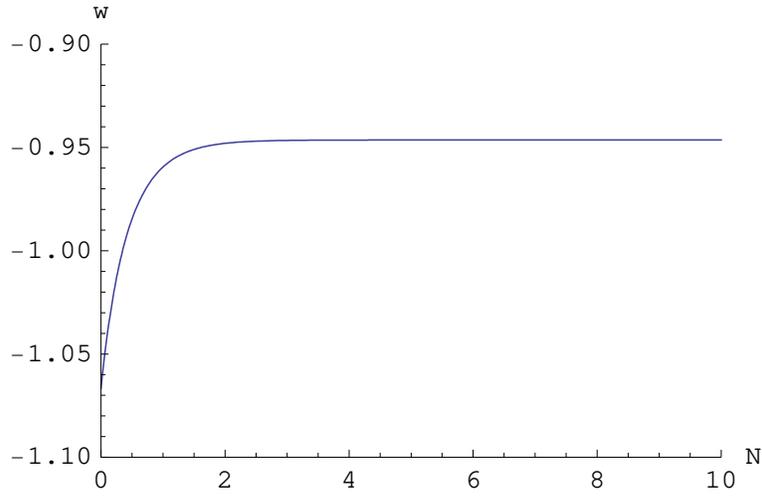}
\caption{\label{Fig3} The evolution of $w$ for total cosmic energy
with $w_x=-1.2$ and $c=0.2$.   }
\end{figure}

Since in LQC the interacting quintessence model has the same
dynamics as that in Einstein cosmology, thereafter   we will only
discuss the case of phantom $w_x<-1$. In Figs.~(\ref{Fig2},
\ref{Fig3}), we plot the numerical results for $c$ and $w_x$
satisfying the conditions $\frac{1+w_x}{w_x}\leq c\leq
\frac{-w_x}{4}$ and $w_x>-2$. Fig.~\ref{Fig2} shows the evolutionary
properties of the universe controlled by the interacting phantom
energy  with $w=-1.2$, $c=0.2$ and different initial conditions.
Apparently the trajectories converge to the same final state
determined by parameter $c$ and $w_x$. Since in the final state
$\Omega_{x}=u_c$ and $\Omega_{m}=1-u_c$, our universe will contain
both dark matter and phantom energy, and the energy ratio between
them approaches a constant.   Fig.~(\ref{Fig3}) shows the
evolutionary curve of the equation of state for total cosmic energy
$w=p_{x}/(\rho_{x}+\rho_{m})$ with $w_x=-1.2$ and $c=0.2$, we find
in the final state  the equation of state is a constant and $w>-1$,
which means that the total energy density decreases with the cosmic
expansion but the universe accelerates forever. Therefore if
$\frac{1+w_x}{w_x}\leq c\leq \frac{-w_x}{4}$ and $w_x>-2$,
regardless the initial conditions,  the universe will enter a final
state with a constant energy ratio between dark energy and dark
matter and accelerate forever.

\begin{figure}[htbp]
\includegraphics[width=10cm]{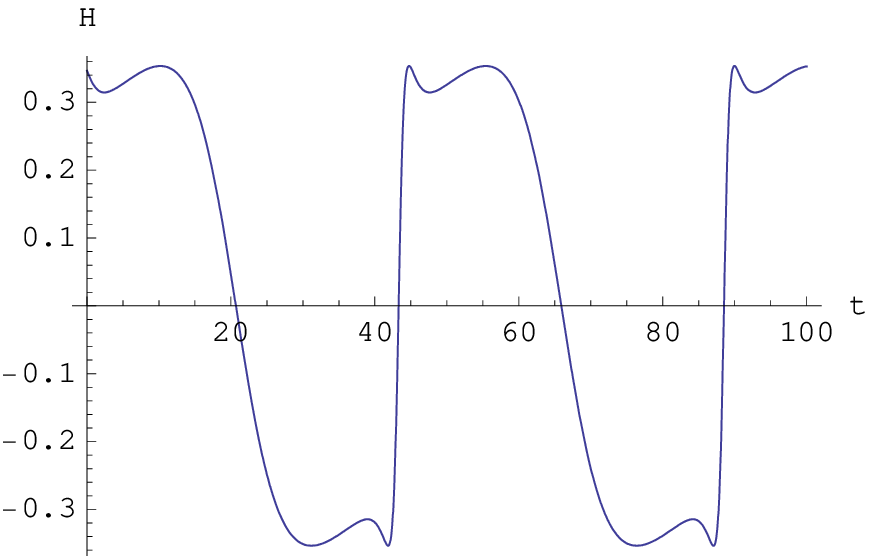}
\caption{\label{Fig4} The evolution of $H$ with time under the
condition of $c<\frac{1+w_x}{w_x}$. Parameters are set as $w=-1.2$,
$c=0.1$ and $\rho_c=1.5$.}
\end{figure}

\begin{figure}[htbp]
\includegraphics[width=10cm]{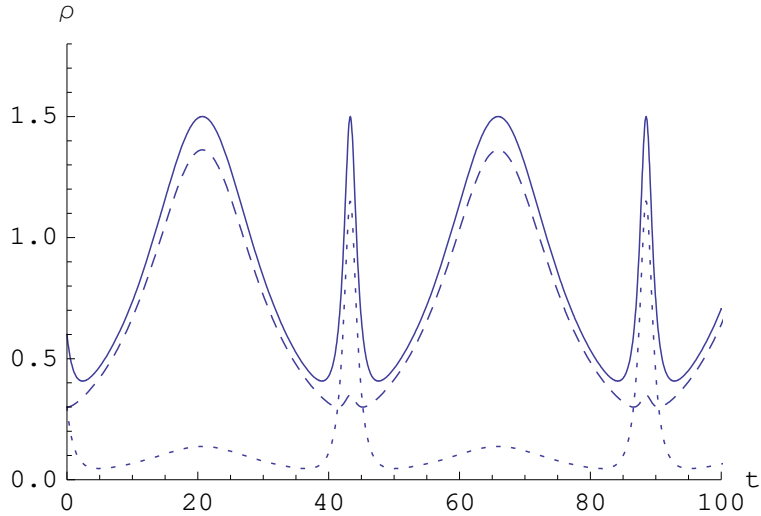}
\caption{\label{Fig5} The evolution of cosmic energy density with
time under the condition of $c<\frac{1+w_x}{w_x}$. The solid, dashed
and dotted curves correspond to $\rho_x+\rho_m$,  $\rho_x$ and
$\rho_m$  respectively. Parameters are set as $w=-1.2$, $c=0.1$ and
$\rho_c=1.5$. Obviously $\rho_x$ triggers the recollapses, while
$\rho_m$ triggers the bounces.}
\end{figure}

\begin{figure}[htbp]
\includegraphics[width=10cm]{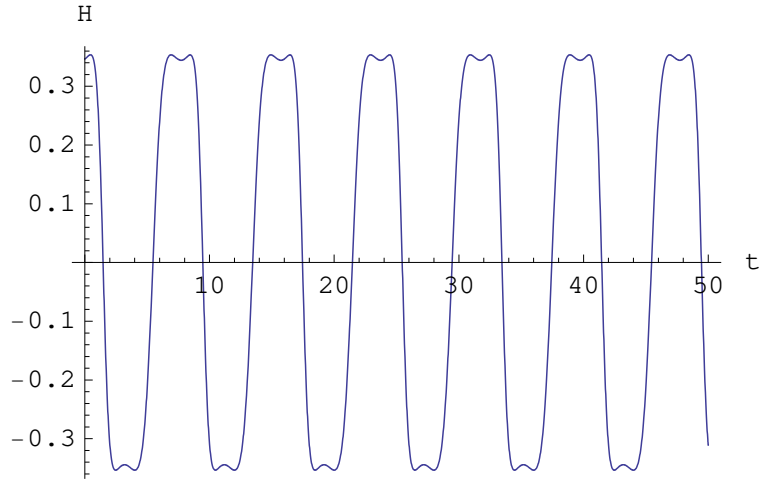}
\caption{\label{Fig6} The evolution of $H$ with time under the
condition of $w_x<-2$. Parameters are set as $w=-2.5$, $c=0.25$ and
$\rho_c=1.5$.}
\end{figure}

\begin{figure}[htbp]
\includegraphics[width=10cm]{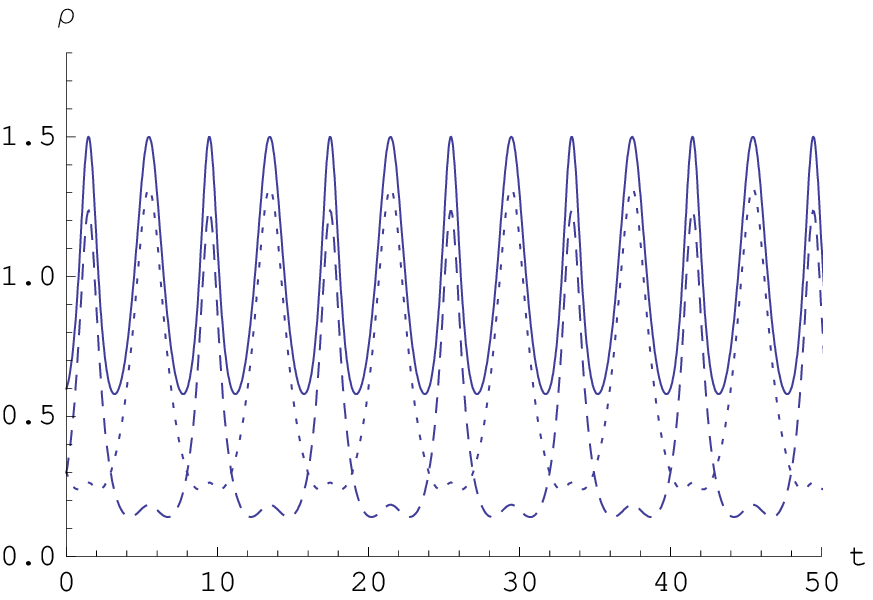}
\caption{\label{Fig7} The evolution of cosmic energy density with
time under the condition of $c>\frac{-w_x}{4}$. The solid, dashed
and dotted curves correspond to $\rho_x+\rho_m$,  $\rho_x$ and
$\rho_m$ respectively. Parameters are set as $w=-2.5$, $c=0.25$ and
$\rho_c=1.5$.}
\end{figure}

In the following we will give the numerical results for the cases
$c<\frac{1+w_x}{w_x}$,  $w_x<-2$  and $c>\frac{-w_x}{4}$ according
to the Eqs.~(\ref{rhox}, \ref{rhom}, \ref{ray}) with $\rho_c=1.5$.
The cases $c<\frac{1+w_x}{w_x}$ and $w_x<-2$  are allowed for the
stable solution in classical Einstein cosmology but ruled out by
loop quantum effect. In Figs.~(\ref{Fig4}, \ref{Fig5}) we plot the
evolutionary curves of $H(t)$ and $\rho(t)$ for the case
$c<\frac{1+w_x}{w_x}$ with $w_x=-1.2$ and $c=0.1$. In
Figs~(\ref{Fig6}, \ref{Fig7}) we give the results for the case
$w_x<-2$ with $w_x=-2.5$ and $c=0.25$. It is easy to find from these
figures that at beginning the phantom energy density increases with
time, which then leads to the increase of total cosmic energy
density. When the total energy density equals to $\rho_c/2$, $H$
takes the maximum value. When $\rho$ reaches its maximum value
$\rho_{max}\sim \rho_c$, $H=0$ and then the universe undergos
contraction until bounce. Therefore the universe will oscillate
forever.

\begin{figure}[htbp]
\includegraphics[width=10cm]{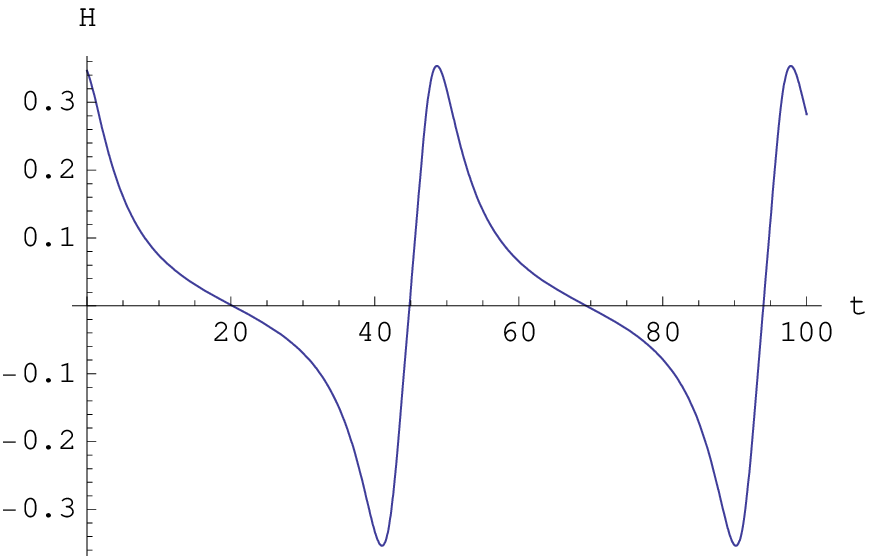}
\caption{\label{Fig8} The evolution of $H$ with time under the
condition of $c>\frac{-w_x}{4}$. Parameters are set as $w=-1.2$,
$c=0.35$ and $\rho_c=1.5$.}
\end{figure}

\begin{figure}[htbp]
\includegraphics[width=10cm]{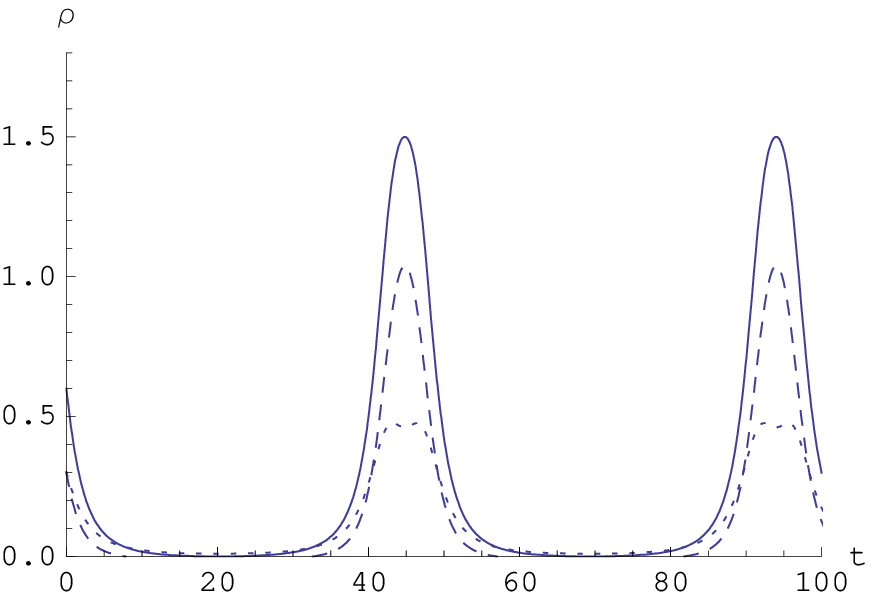}
\caption{\label{Fig9} The evolution of cosmic energy density with
time under the condition of $c>\frac{-w_x}{4}$. The solid, dashed
and dotted curves correspond to $\rho_x+\rho_m$,  $\rho_x$ and
$\rho_m$ respectively. Parameters are set as $w=-1.2$, $c=0.35$ and
$\rho_c=1.5$.}
\end{figure}

Figs~(\ref{Fig8}, \ref{Fig9}) show the results for the case $c >
\frac{-w_x}{4}$ with $w_x=-1.2$ and $c=0.35$ which corresponds to
the case that the critical points do not exist. Comparing these
figures with Figs.~(\ref{Fig4}, \ref{Fig5}, \ref{Fig6}, \ref{Fig7}),
we find, although the universe finally also enters an oscillating
regime, the process is different from that obtained with the
$c<\frac{1+w_x}{w_x}$ or $w_x<-2$.   It is found that in this case
the energy densities  of dark energy and dark matter have the same
evolution with time, and the $H$ changes from positive to negative
(or inverse) when $\rho=0$ or $\rho\approx\rho_c$, while in case
$c<\frac{1+w_x}{w_x}$ or $w_x<-2$,   $H$ changes from positive to
negative (or inverse) only at $\rho\approx\rho_c$

\section{Conclusion}
In this paper the cosmological evolution with the interacting
phantom or quintessence dark energy in loop quantum cosmology is
studied. We consider the case of dark energy with a constant
equation of state $w_x$ and the interaction term with the form
$\Gamma=3Hc(\rho_{x}+\rho_{m})$. It is found that in LQC the dynamic
of interacting quintessence model is the same as that obtained in
classical Einstein cosmological; whereas  for interacting phantom
model, the loop quantum effect reduces significantly the parameter
space, in which the attractor solution exists. In LQC we obtain the
critical point is existent and stable under the conditions
$\frac{1+w_x}{w_x}\leq c\leq \frac{-w_x}{4}$ and $w_x>-2$; however
in classical Einstein cosmology only the condition  $c\leq
\frac{-w_x}{4}$ is required. If the coupling parameter $c$ satisfies
the stable and existent conditions for stable tracking solution, our
universe will enter an ere dominated by both dark energy and dark
matter with a constant energy ratio between them, and accelerate
forever, although the total energy density decreases with cosmic
expansion, otherwise the universe will enter an oscillatory regime.

Recently using the WMAP3~\cite{WMAP3} data Olivares et
al.~\cite{Olivares} obtained that at $1 \sigma$ confidence level
$c\leq 0.0023$. More recently in Ref.~\cite{Wang} by combining the
Gold Sne Ia, BAO and CMB data the authors found that at $1 \sigma$
confidence level $-0.99<w<-0.83$ and $c=0.0057_{-0.0026}^{+0.0030}$,
which show that at  $1 \sigma$ confidence level our universe will
enter a final stable state and can not oscillate. Letting $c=0.0057$
we find in LQC if $-1.006\leq w<-1$ the universe with an interaction
between dark matter and dark energy will accelerate forever; whereas
if $w<-1.006$ it will enter an oscillatory regime. Therefore it is
clear that at $2\sigma$ confidence level the current observations
seem to be unable to predict the late time evolution of our universe
with the interacting dark energy in LQC .

\begin{acknowledgments}
We are grateful to Professor Hongwei Yu for helpful discussions. P.
Wu is partially supported by the National Natural Science Foundation
of China  under Grant No. 10705055,  the Youth Scientific Research
Fund of Hunan Provincial Education Department under Grant No.
07B085, and the Hunan Provincial Natural Science Foundation of China
under Grant No.  08JJ4001. S. N. Zhang  is supported in part by the
Ministry of Education of China, Directional Research Project of the
Chinese Academy of Sciences under project No. KJCX2-YW-T03 and by
the National Natural Science Foundation of China under project Nos.
10521001, 10733010 and 10725313.
\end{acknowledgments}

\end{document}